\documentclass[nofootinbib,reprint,amsmath,amssymb,showkeys, showpacs,aps]{revtex4-1}
\usepackage{graphicx}
\usepackage[T1]{fontenc}
\usepackage{dcolumn}
\usepackage{xcolor,colortbl}
\usepackage{multirow}
\usepackage{bm}
\newcommand{\qm}[1]{``#1''}
\PassOptionsToPackage{breaklinks}{hyperref}
\usepackage{hyperref}
\hypersetup{colorlinks, linkcolor={red},citecolor={blue},urlcolor={blue}}  
\newcommand\ChangeRT[1]{\noalign{\hrule height #1}}

\usepackage{accents}
\usepackage{scalerel}
\usepackage{tabularx,siunitx}
\usepackage{wasysym}

\newcommand{\dd}{{\rm d}}
\usepackage{xurl}

\begin{document}

\title[Static and spherically symmetric wormholes in metric-affine theories of gravity]{ Static and spherically symmetric wormholes in metric-affine theories of gravity}

\author{Vittorio De Falco$^{1,2}$}\email{v.defalco@ssmeridionale.it}
\author{Salvatore Capozziello$^{3,2,1}$}\email{capozziello@unina.it}

\affiliation{$^1$ Scuola Superiore Meridionale,  Largo San Marcellino 10, 80138 Napoli, Italy,\\
$^2$ Istituto Nazionale di Fisica Nucleare, Sezione di Napoli, Complesso Universitario di Monte S. Angelo, Via Cinthia Edificio 6, 80126 Napoli, Italy,\\
$^3$ Universit\`{a} degli studi di Napoli \qm{Federico II}, Dipartimento di Fisica \qm{Ettore Pancini}, Complesso Universitario di Monte S. Angelo, Via Cintia Edificio 6, 80126 Napoli, Italy.}

\date{\today}

\begin{abstract}
We consider static and spherically symmetric wormhole solutions in extended metric-affine theories of gravity supposing that stability and traversability of these objects can be achieved by means of the more geometric degrees of freedom with respect to general relativity. In particular, we concentrate on $f(R)$ metric, $f(T)$ teleparallel, and $f(Q)$ symmetric teleparallel models, where $f$ is a smooth function of curvature, torsion, and non-metricity, respectively. In these extended frameworks, $R,T,$ and $Q$ rule entirely the background geometry without the need to invoke any exotic energy-momentum tensor as matter field source. Assuming that the solution does not present any kind of singularities, we start from the flaring out and null energy conditions to gather together a series of constraints. They allow us to have general indications, which are only necessary but not sufficient and concern only the existence of a throat rather than a global spacetime configuration, to build up then traversable wormholes through a purely geometric approach. The stability cannot be simply assessed via inequalities, because it requires the explicit solution for a detailed analysis.
\end{abstract}

\maketitle
\section{Introduction}
\label{sec:intro}
Wormholes (WHs) are exotic (astrophysical) compact objects configuring as particular and non-trivial topological structures endowed with no horizons and no singularities. Physically, they structure as spacetime-tunnels to connect two isolated regions of the same universe, or even also two completely different universes \cite{Visser1995,Bambi2021}. 

Historically, the concept of WH can be dated back to 1916, few months after Schwarzschild published the first solution to the Einstein field equations, in a paper by Flamm \cite{Flamm1915}. He conjectured the hypothesis of a \qm{\emph{gravitational conduit}}, based on the possibility to create a connection between two regions of the same or different spacetimes via a \emph{black hole} (being an attracting region and configuring thus as the \qm{entrance}) and a \emph{white hole} (being a time reversal of a black hole, which ejects matter, fulfilling thus the role of the \qm{exit}). This line of thinking was refined more mathematically in 1921 by Weyl in relation to his studies on the mass in terms of the electromagnetic field energy \cite{Scholz2012}. 

Flamm's initial ideas were resumed more formally in 1935 by Einstein \& Rosen within the theory of general relativity (GR) \cite{Einstein1935}. However, the terminology \qm{wormhole} was introduced by Misner and Wheeler in 1957, defining such objects as \emph{spaces with muliply-connected topology} \cite{Misner1957}. In 1962 nevertheless, Fuller and Wheeler proved that the Einsten-Rosen WH is unstable, because the light, or any timelike particle, is not able to cross this region \cite{Fuller1962}. The first correct mathematical treatment of WHs was provided independently by Ellis and Bronnikov between 1969 -- 1973 through the \emph{Ellis drainhole model}, representing a static, spherically symmetric, and traversable WH solution to the Einstein vacuum field equations sourced by a scalar field $\phi$, minimally coupled to the geometry of spacetime through a negative coupling polarity (opposite to the standard positive one). In 1988 Morris \& Thorne, revised this description by requiring the presence of \emph{negative energy}. This was achieved via some forms of \emph{exotic matter}, being different from the standard perfect fluids generally sourcing GR field equations \cite{Morris1988}.

In the current scenario, we know that there are two main issues to be taken into account to construct  WH solutions, which are: \emph{traversability} and \emph{energy conditions}. The former is connected to the fact that the WH shortcut occurs through a \emph{minimal surface area}, called WH throat \cite{Hochberg1997a}. Geometrically, it is obtained through the \emph{flaring out condition}, which translates into imposing an inequality involving the shape function and its derivative \cite{Morris1988,Kim2013}. The latter point is more delicate and \emph{strongly depends on the chosen gravity framework}. In GR, we have that employing a standard matter energy-momentum tensor $T_{\mu\nu}^{(sm)}$, the radial tension overcomes the mass-energy density, fulfilling thus the \emph{null energy condition} (NEC), i.e., $T_{\mu\nu}^{(sm)}k^\mu k^\nu\ge0$ where $k^\mu$ is any null vector \cite{Morris1988,Hochberg19981,Hochberg19982,Visser1995}. On the other hand, if we resort to an exotic matter energy-momentum tensor $T_{\mu\nu}^{(em)}$, it violates the NEC and then $T_{\mu\nu}^{(em)}k^\mu k^\nu<0$, which entails negative energy and may be addressed by invoking quantum field theory effects (see e.g., \cite{Hochberg1997b,Bronnikov2013,Calza2018,Garattini2019}). A different situation occurs in extended theories of gravity \cite{CapozzielloRept,Clifton2012}, where it is possible to construct WH solutions with $T_{\mu\nu}^{(sm)}$ without violating the NEC (see e.g., \cite{Harko2013,caplobo1,Duplessis2015,Antoniou2020,Calza2022,Bakopoulos2022,Capozziello:2022zoz}). In all gravity models, the traversability alone is not sufficient to construct a WH, but we need another fundamental ingredient, represented by the \emph{stability}. This implies that the WH throat keeps open for sufficiently long times (see e.g., \cite{Armendariz-Picon2002,Lobo2006,Capozziello2021}).  

In the WH literature, several and different techniques have been exploited to deal with the above-mentioned issues, among which we cite: quantum mechanical-based matter fields \cite{Garattini2019,Parsaei2020} or cosmological fluids \cite{Lukmanova2016,Cataldo2017}; cutting and pasting method to minimize the NEC violation \cite{Garcia2012,Wielgus2020}; thin shell WHs obtained by smoothly gluing together the interior WH solution with the exterior Schwarzschild metric via the Israel junction condition \cite{Garcia2012,Varela2015}; setting the WHs in extended/alternative theories of gravity (see e.g., \cite{Capozziello2012,Capozziello2021,Mustafa2021,Capozziello:2022zoz}), where NEC results to be improved and generalised \cite{caplobo1}.
 
In the realm of gravity frameworks beyond GR, it may be possible to construct WHs in terms of gravity only thanks to the more degrees of freedom. Specifically, in this work we study static and spherically symmetric WH solutions in $f(R)$, $f(T)$, and $f(Q)$ gravity theories in the vacuum, with $f$ being a smooth function of the geometric invariant on which it depends. Therefore, we first a-priori impose the non-singular properties which the WH solution must fulfill. Then, we require the WH traversability by checking the validity of the NEC only in terms of pure gravity. In this setting, the gravitational curvature fluid energy-momentum tensor\footnote{In these extended gravity frameworks beyond GR, we can rearrange the field equations in the vacuum in the GR-like form, namely $G_{\mu\nu}=\chi \mathcal{T}_{\mu\nu}$. $G_{\mu\nu}$ is the Einstein tensor, $\chi$ plays the role of a coupling constant, and $\mathcal{T}_{\mu\nu}$ is the gravitational curvature fluid energy-momentum tensor stemming from the more geometrical degrees of freedom of the underlying theory.} becomes the main responsible for sustaining such peculiar topological structures. Instead, the stability is a more thorny issue, that requires the explicit solution for carrying out further non-trivial calculations. Once all these issues have been checked, we finally ascertain to have determined a WH solution. 

The paper is structured in three parts: first, we recall the properties of a static and spherically symmetric Morris \& Thorne-like WH (Sec. \ref{sec:ss_wormholes}); then we investigate the traversability, together with other non-singularity conditions, in a pure gravity framework within $f(R)$, $f(T)$, and $f(Q)$ theories (Sec. \ref{sec:conditions}); finally in Sec. \ref{sec:end}, we discuss the obtained results, comment on the stability requirement, draw the conclusions, and outline future perspectives. 

\section{Static and spherically\\ symmetric wormholes}
\label{sec:ss_wormholes}
Let us consider WHs in the extended theories of gravity $f(R)$, $f(T)$, and $f(Q)$ in the vacuum (sourced only by gravity), without resorting to any kind of exotic matter energy-momentum tensor. We further impose the \emph{spherical symmetry condition} on the WH solutions. In this respect, it is useful to mention the \emph{Birkhoff theorem}, which would assure that the WH solution is also \emph{static}. It is possible to show that this theorem does not hold in the whole $f(R)$ model, but only in the subcase where $R=R(r)$ (namely $R$ is time-independent) \cite{Faraoni2010,Nzioki2013}. Instead, in the $f(T)$ theory this theorem is applicable if we impose null spin connections (namely the Weitzenb\"{o}ck connection) \cite{Meng2011,Dong2012}. Eventually, in the $f(Q)$ gravity the validity of the Birkhoff theorem has not yet proved, but the situation can be approached similarly to what has been done in the case of the $f(T)$ gravity.

Given these premises, we consider static and spherically symmetric WHs, which are described, in spherical-like coordinates $(t,r,\theta,\varphi)$ and geometrical units $G=c=1$ (independently of the gravity framework which they belong to), by the \emph{Morris \& Thorne-like metric} \cite{Morris1988} 
\begin{equation} \label{eq:MTmetric}
\begin{aligned}
&\dd s^2=-e^{2\Phi(r)}\dd t^2+\frac{\dd r^2}{1-b(r)/r}+r^2\dd\Omega^2,
\end{aligned}
\end{equation}
where $\dd\Omega^2=\dd\theta^2+\sin^2\theta \dd\varphi^2$, $\Phi(r)$ and $b(r)$ are the \emph{redshift and shape functions}, respectively. Equation (\ref{eq:MTmetric}) describes a two-parameter family of metrics depending only on $\Phi(r)$ and $b(r)$, which must be constructed to give a traversable and stable configuration. It is important to note that a WH solution can be assigned either $(i)$ \emph{a-priori}, namely by solving some field equations in a given gravity framework (see e.g., \cite{Lemos2003,Lobo2005,Varela2015,Capozziello2021}), or $(ii)$ \emph{a-posteriori}, meaning that it can be reconstructed via astrophysical methods through the fit of the observational data, once they will be discovred (see e.g., \cite{Cardoso2019,Defalco2020WH,DeFalco2021,DeFalco2021WF,DeFalco2021EF,Defalco2023}).

We follow the definition of a WH (i.e., a self-gravitating system having no horizons and singularities, and being stable and traversable) and require that this object is asymptotically flat, because we want to model isolated gravitational sources. In addition, we consider that the WH solution is symmetric with respect to the WH mouth, which translates in having a symmetric geometry in the two universes. Given these premises, the geometrical properties of metric \eqref{eq:MTmetric} can be stated as follows \cite{Morris1988}:
\begin{enumerate}
\item defined the \emph{WH throat} with $r_{\rm min}$, $\Phi(r)$ is a smooth and everywhere finite function for all $r\ge r_{\rm min}$, with $\Phi(r)\to0$ as $r\to+\infty$. In other words, we require that it is non-singular through all over the two universes, as well as at the WH mouth;
\item $b(r)$ is a smooth function for all $r\ge r_{\rm min}$ with the additional constraints: (1) $b(r)\le r$ (for defining a proper radial distance $l$); (2) the presence of the WH throat, defined as the minimum radius, implies $r_{\rm min}=b_0$ and $b(r_{\rm min})=b_0$; (3) $b'(r)<b(r)/r$ near the WH throat (\emph{flaring out condition}), where from now on the prime stays for the derivative with respect to $r$; (4) $b(r)/r\to0$ as $r\to+\infty$.
\end{enumerate}

The WH mass $M$ is defined, according to the Arnowitt, Deser, Misner (ADM) formalism, as the total mass of the system contained in the whole spacetime \cite{Visser1995}, i.e., 
\begin{equation} \label{eq:ADMmass}
M\equiv \lim_{r\to+\infty}m(r)=\frac{b_0}{2}+4\pi\int^{\infty}_{b_0}\rho(x)x^2 dx,
\end{equation}
where $\rho(r)$ represents the mass-energy density. This definition is also known as the \emph{Morris-Thorn mass function}, which can be easily derived by employing Eq. (14) in Ref. \cite{Morris1988}. The gravity theory enters in the function $\rho(r)$, since we can generally recast the field equations (in whatever gravity framework) in a GR-like form. Another important remark is that Eq. \eqref{eq:ADMmass} describes the mass in the whole zones between the WH throat, albeit the integral extends only over one region. This apparent contradiction is resolved by considering the symmetry existing between the two universes. Beside the mass, the WH solution (i.e., $\Phi(r)$, and $b(r)$) can also depend on other $n$ parameters $q_1,\cdots,q_n$, stemming from the underlying gravity framework (see e.g, Refs. \cite{DeFalco2021,DeFalco2021EF,Defalco2023}). 

\section{Traversable wormhole solutions}
\label{sec:conditions}
In this section, we analyse the WH solutions \eqref{eq:MTmetric} in the vacuum field equations in $f(R)$ (Sec. \ref{sec:f(R)_theory}), $f(T)$ (Sec. \ref{sec:f(T)_theory}), and $f(Q)$ (Sec. \ref{sec:f(Q)_theory}) theories of gravity. 

In each one of the above-mentioned gravity frameworks, we impose the validity of the NEC near the WH throat. Without loss of generality, we can study the problem in the equatorial plane and we can set $\theta=\pi/2$ in Eq. \eqref{eq:MTmetric}. We use the symbol $\nabla_\alpha$ to denote the covariant derivative, which acts on a generic $(1,1)$ tensor $A^\mu_{\ \nu}$ as 
\begin{equation}
\nabla_\alpha A^\mu_{\ \nu}=\partial_\alpha A^\mu_{\ \nu}-\Gamma^\lambda_{\ \nu\alpha} A^\mu_{\ \lambda} + \Gamma^\mu_{\ \lambda\alpha} A^\lambda_{\ \nu},  
\end{equation}
where $\Gamma^\mu_{\ \alpha\beta}$ is the affine connection in the considered theory.  The field equations descend from the action
\begin{equation} \label{eq:action}
S:=\int f(X)\sqrt{-g}\ \dd^4x,
\end{equation} 
where $f$ is a smooth function of $X=R$, or $T$, or $Q$, and $g$ is the determinant of the metric \eqref{eq:MTmetric}. In the particular case where $f(X)=x_0\ X+\Lambda$ with $x_0$ and $\Lambda$ being constants, the three theories are all \emph{dynamically equivalent to GR}, giving rise to the so-called  \emph{geometric trinity of gravity} \cite{Beltran2019,Capozziello2021GT,Capozziello:2022zzh}. By minimizing the action with respect to the metric tensor for $f(R)$ and $f(Q)$, and the tetrad field for $f(T)$, we obtain the related field equations, which can be all recast in the GR-like form as follows \cite{Faraoni2010,Atazadeh2013,DAmbrosio2022}
\begin{equation} \label{eq:field_equations}
G_{\mu\nu}=\frac{1}{\dot{f}(X)}T_{\mu\nu}^{\rm (curv)}.
\end{equation}
Here, $G_{\mu\nu}$ is the Riemannian Einstein tensor, $\dot{f}(X):=\partial_X f(X)$ and, more in general,  over-dots represent the derivatives of the function $f(X)$ with respect to $X$, and $T_{\mu\nu}^{\rm (curv)}$ is the \emph{(gravitational) curvature fluid energy-momentum tensor}, which assumes different forms in the three above-cited gravity frameworks. It is worth noticing that $1/\dot{f}(X)$ plays the role of an effective gravitational coupling. Imposing that the graviton is not a ghost, it is physically consistent with the assumption $\dot{f}(X)>0$ (see Ref. \cite{Sotiriou2010}, and references therein).

Therefore, the WH traversability is related to the form of $T_{\mu\nu}^{\rm (curv)}$. Pursuing the approach devised by Morris \& Thorne, we consider the following tetrad field \cite{Morris1988}
\begin{equation} \label{eq:TETRAD}
e^A_{\ \alpha}={\rm diag}(\sqrt{-g_{tt}},\ \sqrt{g_{rr}}, \sqrt{g_{\theta\theta}},\ \sqrt{g_{\varphi\varphi}}),
\end{equation} 
where capital Latin indices refers to the tangent Minkowski spacetime, whereas Greek ones are the indices on the manifold. This allows to define the (proper) total mass-energy density $-\rho(r)$, and the (proper) radial and tangential pressures $p_r(r)$ and $p_t(r)$, respectively. Their expressions are given by the following transformations
\begin{subequations} \label{eq:components_curvature_tensor}
\begin{align}
-\rho(r)&:=T_{\hat t \hat t}^{\rm (curv)}\equiv e^A_{\ t}e^B_{\ t}T_{AB}^{\rm (curv)}= \frac{1}{g_{tt}}T_{tt}^{\rm (curv)},\\
p_r(r)&:=T_{\hat r \hat r}^{\rm (curv)}\equiv e^A_{\ r}e^B_{\ r}T_{AB}^{\rm (curv)}= \frac{1}{g_{rr}}T_{rr}^{\rm (curv)},\\
p_t(r)&:=T_{\hat \varphi \hat \varphi}^{\rm (curv)}\equiv e^A_{\ \varphi}e^B_{\ \varphi}T_{AB}^{\rm (curv)}= \frac{1}{g_{\varphi\varphi}}T_{\varphi\varphi}^{\rm (curv)}.
\end{align}
\end{subequations}
Therefore, we must impose in each gravity environment the validity of the NEC, namely \cite{caplobo1}
\begin{subequations} \label{eq:NEC}
\begin{align}
&(\rho+p_r)\ge0,  \label{eq:NEC_r}\\
&(\rho+p_t)\ge0. \label{eq:NEC_phi}
\end{align}
\end{subequations}

\subsection{$f(R)$ metric gravity}
\label{sec:f(R)_theory}
We first briefly describe the gravity framework (Sec. \ref{sec:f(R)_geometry}), and then analyse the NEC conditions \eqref{eq:NEC}  providing thus constraints on both the theory itself and the metric components (Sec. \ref{sec:f(R)_WH_cond}).

\subsubsection{Geometric structure and field equations}
\label{sec:f(R)_geometry}
The (metric) $f(R)$ gravity  is constructed upon the symmetric metric tensor $g_{\mu\nu}$ \eqref{eq:MTmetric}, from which it is then possible to determine the other geometric quantities. 

Based on the \emph{metric compatibility condition} $\nabla_\alpha g_{\mu\nu}=0$, we can  define the Levi-Civita connection 
\begin{equation} \label{eq:LC_connection}
\Gamma^\alpha_{\ \mu\nu}:=\frac{1}{2}g^{\alpha\rho}\biggr{(}\partial_\mu g_{\rho\nu}+\partial_\nu g_{\mu\rho}-\partial_\rho g_{\mu\nu}\biggr{)},
\end{equation} 
which then allows to introduce the Riemann tensor
\begin{equation}
R^\alpha_{\ \beta\mu\nu}:=\partial_\mu \Gamma^\alpha_{\ \beta\nu}-\partial_\nu \Gamma^\alpha_{\ \beta\mu}+\Gamma^\alpha_{\ \lambda\mu}\Gamma^\lambda_{\ \beta\nu}-\Gamma^\alpha_{\ \lambda\nu}\Gamma^\lambda_{\ \beta\mu},
\end{equation}
as well as the Ricci tensor and scalar curvature
\begin{subequations}
\begin{align}
R_{\alpha\beta}&:=R^\mu_{\ \alpha\mu\beta},\label{eq:Ricci_tensor}\\
R&:=g^{\alpha\beta}R_{\alpha\beta}. \label{eq:scalr_curvature}
\end{align}
\end{subequations}
$T_{\mu\nu}^{\rm (curv)}$ in the field equations \eqref{eq:field_equations} reads as \cite{Faraoni2010,Capozziello:2012uv}
\begin{align}
T_{\mu\nu}^{\rm (curv)}&:=\frac{1}{2}g_{\mu\nu}\biggr{[}f(R)-R\dot{f}(R)\biggr{]}+\nabla_\mu\nabla_\nu \dot{f}(R)\notag\\
&-g_{\mu\nu}\Box \dot{f}(R),
\end{align}
where $\Box:=g^{\mu\nu}\nabla_\mu \nabla_\nu$ is the curved d'Alembert operator.

\subsubsection{Conditions for traversable wormholes in $f(R)$ gravity}
\label{sec:f(R)_WH_cond}
Starting from the assumptions $\dot{f}(R)>0$ and $R(r)>0$, we  work out Eq. \eqref{eq:NEC_r}, which entails
\begin{equation} \label{eq:NEC1}
-\frac{\nabla_t\nabla_t \dot{f}(R)}{g_{tt}}+\frac{\nabla_r\nabla_r \dot{f}(R)}{g_{rr}}\ge0.
\end{equation}
Exploiting the following identity
\begin{equation}
\nabla_\mu \nabla_\nu \dot{f}(R)=\partial_\mu\partial_\nu \dot{f}(R)-\partial_r  \dot{f}(R)\Gamma^r_{\ \mu\nu},
\end{equation}
where $\partial_\alpha \dot{f}(R) \Gamma^\alpha_{\ \mu\nu}=\partial_r \dot{f}(R) \Gamma^r_{\ \mu\nu}$ since the metric components depend only on the radial coordinate $r$. 
Therefore, Eq. \eqref{eq:NEC1} becomes 
\begin{align} \label{eq:NEC2}
&[\dot{f}(R)]'\frac{\Gamma^r_{\ tt}}{g_{tt}}+[\dot{f}(R)]''\frac{1}{g_{rr}}-[\dot{f}(R)]'\frac{\Gamma^r_{\ rr}}{g_{rr}}\notag\\
=&R'(r)\biggr{\{}\ddot{f}(R)\biggr{[}\frac{\Gamma^r_{\ tt}}{g_{tt}}+\frac{R''(r)}{g_{rr}R'(r)}-\frac{\Gamma^r_{\ rr}}{g_{rr}}
\biggr{]}\notag\\
&\quad+\frac{\dddot{f}(R) R'(r)}{g_{rr}}
\biggr{\}}\ge0\,,
\end{align}
where the prime is the derivative with respect to the radial coordinate. The affine connections appearing in Eq. \eqref{eq:NEC2} have the following explicit expressions (cf. Eq. \eqref{eq:LC_connection})
\begin{subequations}
\begin{align}
\Gamma^r_{\ tt}&=e^{2\Phi(r)}\Phi'(r)\left(1-\frac{b(r)}{r}\right),\\
\Gamma^r_{\ rr}&=-\frac{b(r)-b'(r)r}{2r^2}\left(1-\frac{b(r)}{r}\right)^{-1}.
\end{align}
\end{subequations}
Therefore, Eq. \eqref{eq:NEC2} can be written as
\begin{align} \label{eq:NEC2bis}
&R'(r)\biggr{\{}\ddot{f}(R)\biggr{[}\left(1-\frac{b(r)}{r}\right)\left(-\Phi'(r)+\frac{R''(r)}{R'(r)}\right)\notag\\
&+\frac{b(r)-b'(r)r}{2r^2}\biggr{]}+\dddot{f}(R) R'(r)\left(1-\frac{b(r)}{r}\right)\biggr{\}}\ge0.
\end{align}
Evaluating this expression at the WH throat, we obtain
\begin{align}
&[R'(r)\ddot{f}(R)]_{r=b_0}\biggr{[}\frac{b(r_0)-b'(r_0)r_0}{2r_0^2}\biggr{]}\ge0.
\end{align}
The term in the square bracket can be neglected using the flaring out condition (see Sec. \ref{sec:ss_wormholes}), which leads to
\begin{align} \label{eq:constraint1}
[R'(r)\ddot{f}(R)]_{r=b_0}\ge0 \quad \Rightarrow\quad R'(r)\ddot{f}(R)\ge0,
\end{align}
where the extension to general $r$ can be done because we suppose that $R(r)>0$ and therefore the sign of $\ddot{f}(R)$ is the same at $b_0$ and in a generic $r$, therefore the product $R'(r)\ddot{f}(R)$ keeps always the same sign. 

Using the above result, we have that Eq. \eqref{eq:NEC2bis} gives a further constraint, i.e., 
\begin{align} \label{eq:constraint1bis}
\dddot{f}(R)\ge -\frac{\ddot{f}(R)}{R'(r)}\left[\frac{b(r)-b'(r)r}{2r(r-b(r))}+\Phi'(r)-\frac{R''(r)}{R'(r)}\right].
\end{align}

Let us now  focus on the other NEC \eqref{eq:NEC_phi}. Along the line followed above, using
\begin{equation} 
\Gamma^r_{\ \varphi\varphi}=-(r-b(r)),
\end{equation} 
it leads to
\begin{equation} \label{eq:NEC3}
R'(r)\ddot{f}(R)\left(1-\frac{b(r)}{r}\right)\biggr{(}\frac{1}{r}-\Phi'(r)\biggr{)}\ge0.
\end{equation}
As it can be promptly checked, if Eq. \eqref{eq:NEC3} is evaluated at the WH throat, it does not give significant information. However, using Eq. \eqref{eq:constraint1},  we obtain
\begin{equation}
\Phi'(r)\le \frac{1}{r}.
\end{equation}
Now, we take into account the initial constraint $R(r)>0$. The scalar curvature \eqref{eq:scalr_curvature}, related to the metric \eqref{eq:MTmetric}, is
\begin{align}
R(r)&=\frac{1}{r^2}\biggr{\{}b'(r) \biggr{(}r \Phi '(r)+2\biggr{)}-\biggr{(}4 r-3 b(r)\biggr{)} \Phi '(r)\notag\\
&-2 r \biggr{(}r-b(r)\biggr{)} \biggr{(}\Phi ''(r)+\Phi '(r)^2\biggr{)}\biggr{\}}.
\end{align}
Employing $R(r)>0$ and the conditions assumed on the shape function (see Sec. \ref{sec:conditions}), we obtain
\begin{align} \label{eq:inequality1}
&\frac{1}{r \Phi '(r)+2}\biggr{[}2 r \biggr{(}r-b(r)\biggr{)} \biggr{(}\Phi ''(r)+\Phi '(r)^2\biggr{)}\notag\\
&+\biggr{(}4 r-3 b(r)\biggr{)} \Phi '(r)\biggr{]}<b'(r)<\frac{b(r)}{r}\le1.
\end{align}
Evaluating the above expression at the WH throat and using again the flaring out condition (see Sec. \ref{sec:conditions}), we derive the following constraint on $b'(b_0)$
\begin{equation}
\frac{b_0\Phi'(b_0)}{2+b_0\Phi'(b_0)}<b'(b_0)<1.
\end{equation}
Furthermore, considering the terms on the most left hand side of the inequality in Eq. \eqref{eq:inequality1} and imposing that they are minor than $b(r)/r$, it is possible to extract another inequality, which constraints $\Phi'(r)$, namely
\begin{align} \label{eq:inequality2}
\Phi_1(r)<\Phi'(r)<\Phi_2(r)\le \frac{1}{r},
\end{align}
where
\begin{subequations} \label{eq:SYMBOLS}
\begin{align} 
\Phi_1(r)&=-\frac{r(r - b(r))+\sqrt{\Delta(r)}}{r^2 (r-b(r))},\\
\Phi_2(r)&=-\frac{r(r - b(r))-\sqrt{\Delta(r)}}{r^2 (r-b(r))},\\
\Delta(r)&=r^3 (r-b(r)) \biggr{[}-r (r-b(r)) \Phi ''(r)+1\biggr{]}.
\end{align}
\end{subequations}
Imposing that $\Delta(r)>0$, we have this other inequality 
\begin{align} \label{eq:inequality3}
\Phi ''(r)<\frac{1}{r (r-b(r))}.
\end{align}
The condition $\Phi_2(r)\le 1/r$ gives
\begin{equation} \label{eq:inequality4}
b(r)\le r\left[\frac{r^2 \Phi ''(r)+3}{r^2 \Phi ''(r)+4}\right]\le r.
\end{equation}
To resume, we have obtained the following constraints \eqref{eq:constraint1}, \eqref{eq:constraint1bis}, \eqref{eq:inequality1}, \eqref{eq:inequality2}, \eqref{eq:inequality3}, and \eqref{eq:inequality4}, which entail restrictions on theories of gravity and metric components apt to describe WH solutions subject only to pure gravity.  

We highlight an important aspect of the WH solutions in $f(R)$ gravity based on Refs. \cite{Bronnikov2007,Bronnikov2010}, where the authors provides a no-go theorem. It claims that \emph{the nonexistence for static and spherically symmetric WHs in the absence of any exotic matter violating the NEC within the scalar-tensor theories, hence also in the $f(R)$ gravity.} Given these premises, it follows that a throat is possible without invoking exotic matter, but a WH as a global configuration seems to be not possible under reasonable and generic conditions. However, we stress that the inequalities we have determined are only necessary, but not sufficient to declare the existence of WHs.
%To conclude, we could find some viable WH configurations, but they must be scrupulously checked in view of the above-stated theorem, which provides a serious warning on the existence of WH solutions in the $f(R)$ gravity theory.}

\subsection{$f(T)$ teleparallel gravity}
\label{sec:f(T)_theory}
This section describes the $f(T)$ gravity theory (Sec. \ref{sec:f(T)_geometry}), and then analyse the NEC Eqs. \eqref{eq:NEC} to determine the constraints for traversable WHs (Sec. \ref{sec:f(T)_WH_cond}).

\subsubsection{Geometrical structure and field equations}
\label{sec:f(T)_geometry}
In the extended  teleparallel theory of gravity, the fundamental geometric objects are represented by the tetrad field $h^A_{\ \mu}$ and the spin connection $\omega^A_{\ B\mu}$ \cite{Cai2016,Martin2019}. Tetrads solder the metric $g_{\mu\nu}$ with the Minkowoski metric $\eta_{AB}$ through the following equations
\begin{equation}
\eta_{AB}=h_A^{\\ \mu}h_B^{\\ \nu}g_{\mu\nu},\qquad g_{\mu\nu}=h^A_{\\ \mu}h^B_{\\ \nu}\eta_{AB}.
\end{equation}
Tetrads respect also the orthonormality conditions
\begin{equation}
h^A_{\ \mu}h_B^{\ \mu}=\delta^A_B,\qquad h^A_{\ \mu}h_A^{\ \nu}=\delta^\nu_\mu.
\end{equation}
The affine connection is curvature-less and metric compatible, whose explicit expression is given by \cite{Cai2016,Martin2019}
\begin{equation} \label{eq:f(T)_connection}
\Gamma^\alpha_{\ \mu\nu}:=h_A^{\ \ \alpha}\left(\partial_\mu h^A_{\ \nu}+\omega^A_{\ B\mu} h^B_{\ \nu}\right).
\end{equation}
In this extended framework, we have to stress that it is not possible to exploit the Weitzenb\"ock connection and a generic tetrad $h^A_{\ \mu}$, without  obtaining meaningless results. In fact, chosen a tetrad $h^A_{\ \mu}$, we must associate to it the related spin connection $\omega^A_{\ B\mu}$, entrusted to describe the inertial effects inside the tetrad (see Ref. \cite{Martin2019}, and discussions therein). In other words, we must always consider the right couple $\left\{h^A_{\ \mu},\omega^A_{\ B\mu}\right\}$. 

The torsion tensor coincides with the antisymmetric part of the affine connection \eqref{eq:f(T)_connection}, namely \cite{Cai2016,Martin2019}\footnote{We use the notation $A_{[\mu\nu]}=A_{\mu\nu}-A_{\nu\mu}$ and $A_{(\mu\nu)}=A_{\mu\nu}+A_{\nu\mu}$.}
\begin{align}
&T^\alpha_{\ \mu\nu}:=\Gamma^\alpha_{\ [\nu\mu]}\notag\\
&=-h^\alpha_{\ A}\biggr{(}\partial_\mu h^A_{\ \nu}-\partial_\nu h^A_{\ \mu}+\omega^A_{\ B\mu}h^B_{\ \nu}-\omega^A_{\ B\nu}h^B_{\ \mu}\biggr{)}.
\end{align}
We then introduce the contortion tensor \cite{Cai2016,Martin2019}
\begin{equation}
K^\alpha_{\ \mu\nu}:=\frac{1}{2}\left(T_\mu{}^\alpha_{\ \nu}+T_\nu{}^\alpha_{\ \mu}-T^\alpha_{\ \mu\nu}\right),
\end{equation}
as well as the superpotential 
\begin{equation}
S_A^{\ \mu\nu}:=K^{\mu\nu}{}_{A}+h_A^{\ \mu} T^{\beta\nu}{}_\beta-h^{\ \nu}_A T^{\beta\mu}{}_\beta.
\end{equation}
We are now able to define the torsion scalar \cite{Cai2016,Martin2019}
\begin{equation}
T:=\frac{1}{2}T^\alpha_{\ \mu\nu}S_\alpha^{\ \mu\nu}.
\end{equation}
We have all the ingredients to finally write the field equations \eqref{eq:field_equations}, where $T^{\rm (curv)}_{\mu\nu}$ is given by \cite{Atazadeh2013,Martin2019}
\begin{equation}
T^{\rm (curv)}_{\mu\nu}=-\frac{1}{2}g_{\mu\nu}\biggr{[}f(T)-\dot{f}(T)T\biggr{]}-\ddot{f}(T)S_{\nu\mu}{}^\rho\nabla_\rho T.
\end{equation}
The spin connection satisfies the field equations \cite{Martin2019}
\begin{equation} \label{eq:spin_cconnection_field_equations}
\ddot{f}(T)S_{[AB]}{}^\nu\partial_\nu T=0,
\end{equation}
which can be equivalently obtained by varying the action \eqref{eq:action} with respect to the spin connection. 

\subsubsection{Conditions for traversable wormholes in $f(T)$ gravity}
\label{sec:f(T)_WH_cond}
We first set the tetrad \eqref{eq:TETRAD}, i.e., $h^A_{\ \alpha}=e^A_{\ \alpha}$, whose related spin connection is given by \cite{Martin2019}
\begin{subequations}
\begin{align}
%t
\omega^{\hat t}_{\ A\mu}&=\begin{pmatrix}
0 & 0 & 0 & 0 \\
0 & 0 & -1 & 0 \\
0 & 1 &  & -\sin\theta \\
0 & 0 & \sin\theta & 0 \\
\end{pmatrix},\\
%r
\omega^{\hat r}_{\ A\mu}&=\begin{pmatrix}
0 & 0 & 1 & 0 \\
0 & 0 & 0 & 0 \\
-1 & 0 & 0 & -\cos\theta \\
0 & 0 & \cos\theta & 0 \\
\end{pmatrix},\\
%\theta
\omega^{\hat \theta}_{\ A\mu}&=\begin{pmatrix}
0 & 0 & 0 & \sin\theta \\
0 & 0 & 0 & \cos\theta \\
0 & 0 & 0 & 0 \\
-\sin\theta & -\cos\theta & 0 & 0 \\
\end{pmatrix},\\
%\phi
\omega^{\hat \varphi}_{\ A\mu}&=\begin{pmatrix}
0 & 0 & 0 & \sin^2\theta \\
0 & 0 & 0 & \frac{\sin(2\theta)}{2} \\
0 & 0 & 0 & 0 \\
-\sin^2\theta & -\frac{\sin(2\theta)}{2} & 0 & 0 \\
\end{pmatrix},
\end{align}
\end{subequations}
where the overhat index means that it is evaluated in the tetrad frame (corresponding to capital Latin indices).

Assuming $\dot{f}(T)>0$ and $T(r)>0$, we then calculate the NEC \eqref{eq:NEC_r}, which is given by
\begin{align}
&\frac{\ddot{f}(T)S_{tt}{}^\rho\nabla_\rho T}{g_{tt}}-\frac{\ddot{f}(T)S_{rr}{}^\rho \nabla_\rho T}{g_{rr}}\notag\\
=&\frac{\ddot{f}(T)S_{tt}^{\ \ r}\ T'(r)}{g_{tt}}\ge0,
\end{align}
Exploiting the following expression
\begin{equation}
S_{tt}^{\ \ r}=-\frac{2}{r} \left(1-\frac{b(r)}{r}\right) e^{2 \Phi (r)},
\end{equation}
we obtain
\begin{align}
&\ddot{f}(T) T'(r)\frac{2}{r} \left(1-\frac{b(r)}{r}\right)\ge0.
\end{align}
Evaluating this inequality at the WH throat, it gives no significant information. Instead, in general we have
\begin{align} \label{eq:ine1}
&\ddot{f}(T) T'(r)\ge0,
\end{align}
which is valid for all $r$ in the WH solution's domain. 

Moving on the NEC \eqref{eq:NEC_phi} and using
\begin{equation}
S_{\varphi\varphi}{}^r=r^2\left(1-\frac{b(r)}{r}\right) \left(\frac{1}{r}+\Phi'(r)\right),
\end{equation}
we obtain
\begin{align}
\ddot{f}(T)T'(r)\left(1-\frac{b(r)}{r}\right) \left(\frac{1}{r}-\Phi'(r)\right)\ge0.
\end{align}
This inequality implies (cf. Eq. \eqref{eq:ine1})
\begin{align} \label{eq:ine1bis}
\Phi'(r)\le \frac{1}{r}.
\end{align}

The expression of $T(r)$ is
\begin{equation}
T(r)=-\frac{4(r-b(r)) \left(1+2 r \Phi '(r)\right)}{r^3}
\end{equation}
From $T(r)>0$,  we deduce (cf. Eq. \eqref{eq:ine1})
\begin{equation} \label{eq:ine2}
\Phi '(r)< -\frac{1}{2 r}<\frac{1}{r}.
\end{equation}
Therefore, we have found the constraints \eqref{eq:ine1} and \eqref{eq:ine2} for having a WH solution in pure $f(T)$ gravity.

\subsection{$f(Q)$ symmetric teleparallel gravity}
\label{sec:f(Q)_theory}
We first recall the basic definitions of the $f(Q)$ gravity theory (see Sec. \ref{sec:f(Q)_geometry}), and then calculate the NEC Eqs. \eqref{eq:NEC}, where we then derive the related constraints by making us of pure gravity only (see Sec. \ref{sec:f(Q)_WH_cond}).

\subsubsection{Geometrical structure and field equations}
\label{sec:f(Q)_geometry}
In the extended symmetric teleparallel theories of gravity, the fundamental geometrical objects are represented by the metric tensor $g_{\mu\nu}$ and the affine connection $\Gamma^\alpha_{\ \mu\nu}$. They are deputed to describe two independent concepts. The former is ascribed to rule the casual structure, whereas the latter defines the geodesic scaffold. In this framework, they are not required to coincide as in GR. This theory is characterised by vanishing curvature and torsion. The gravitational effects are expressed in terms of the following non-metricity tensor 
\begin{align}
Q_{\alpha \mu\nu}&:=\nabla_\alpha g_{\mu\nu}=\partial_\alpha g_{\mu\nu}-\Gamma^\lambda_{\alpha(\mu}g_{\nu)\lambda}.
\end{align}
Then, we can introduce the non-metricity scalar 
\begin{align}
Q&:=-\frac{1}{4} Q_{\alpha\mu\nu}Q^{\alpha\mu\nu}+\frac{1}{2}Q_{\alpha\mu\nu}Q^{\alpha\nu\mu}+\frac{1}{4}Q_\alpha Q^\alpha\notag\\
&-\frac{1}{2}Q_\alpha \bar{Q}^\alpha,
\end{align}
where
\begin{equation}
Q_{\alpha}=Q_{\alpha\nu}^{\ \ \nu},\qquad \bar{Q}_\alpha=Q^\nu_{\ \nu\alpha}.
\end{equation}
Defined the non-metricity conjugate
\begin{align}
P^\alpha_{\ \mu\nu}&:=-\frac{1}{4}Q^\alpha_{\ \mu\nu}+\frac{1}{4}Q_{(\mu}{}^\alpha_{\ \nu)}+\frac{1}{4}g_{\mu\nu}Q^\alpha\\
&-\frac{1}{4}\left[g_{\mu\nu}\bar{Q}^\alpha+\frac{1}{2}\delta^\alpha_{(\mu}Q_{\nu)}\right],
\end{align}
we have the field Eqs. \eqref{eq:field_equations}, where $T^{\rm (curv)}_{\mu\nu}$ is \cite{DAmbrosio2022}
\begin{equation}
T^{\rm (curv)}_{\mu\nu}=\frac{1}{2}g_{\mu\nu}\biggr{[}f(Q)-\dot{f}(Q)Q\biggr{]}-2\ddot{f}(Q)P^\alpha_{\ \mu\nu}\partial_\alpha Q.
\end{equation}
Varying the action \eqref{eq:action} with respect to the affine connection, we obtain the related field equations 
\begin{equation}
\nabla_\mu \nabla_\nu \left(\sqrt{-g}\dot{f}(Q)P^{\mu\nu}{}_\alpha\right)=0.
\end{equation}

\subsubsection{Conditions for traversable wormholes in $f(Q)$ gravity}
\label{sec:f(Q)_WH_cond}
We assume $\dot{f}(Q)>0$ and $Q(r)>0$, and then we analyse the NEC \eqref{eq:NEC_r}, where we have
\begin{align} \label{eq:inequality_1}
&\frac{2\ddot{f}(Q)P^\alpha_{\ tt}\partial_\alpha Q}{g_{tt}}-\frac{2\ddot{f}(Q)P^\alpha_{\ rr}\partial_\alpha Q}{g_{rr}}\notag\\
&=2\ddot{f}(Q)Q'(r)\left[\frac{P^r_{\ tt}}{g_{tt}}-\frac{P^r_{\ rr}}{g_{rr}}\right]\notag\\
&=-2\ddot{f}(Q)Q'(r)\left[\frac{P^r_{\ tt}}{e^{2\Phi(r)}}+P^r_{\ rr}\left(1-\frac{b(r)}{r}\right)\right]\ge0.
\end{align}
For a static and spherically symmetric solution, we know that the affine connection can be written in the most general form as (see solution set 2 in Ref. \cite{DAmbrosio2022}, for details)
\begin{subequations}
\begin{align}
%t
\Gamma^t_{\ \mu\nu}&=\begin{pmatrix}
\mathcal{A}(r) & \mathcal{B}(r) & 0 & 0 \\
 \mathcal{B}(r) & \Gamma^t_{\ rr}(r) & 0 & 0 \\
0 & 0 & \Gamma^t_{\ \theta\theta}(r) & 0 \\
0 & 0 & 0 & \Gamma^t_{\ \theta\theta}(r)\sin^2\theta \\
\end{pmatrix},\\
%r
\Gamma^r_{\ \mu\nu}&=\begin{pmatrix}
 \mathcal{C}(r) & \mathcal{D}(r) & 0 & 0 \\
 \mathcal{D}(r) & \Gamma^r_{\ rr}(r) & 0 & 0 \\
 0 & 0 & \Gamma^r_{\ \theta\theta}(r) & 0 \\
 0 & 0 & 0 & \Gamma^r_{\ \theta\theta}(r)\sin^2\theta 
\end{pmatrix},\\
%\theta
\Gamma^\theta_{\ \mu\nu}&=\begin{pmatrix}
 0 & 0 & c & 0 \\
 0 & 0 & \mathcal{E}(r) & 0 \\
 c & \mathcal{E}(r) & 0 & 0 \\
 0 & 0 & 0 & -\cos\theta\sin\theta 
\end{pmatrix},\\
%\theta
\Gamma^\varphi_{\ \mu\nu}&=\begin{pmatrix}
 0 & 0 & 0 & c \\
 0 & 0 & 0 & \mathcal{E}(r)\\
 0 & 0 & 0 & \cot\theta \\
 c & \mathcal{E}(r)& \cot\theta & 0 \\
\end{pmatrix},
\end{align}
\end{subequations}
where
\begin{subequations}
\begin{align}
\mathcal{A}(r)&=-c+k-c(2c-k)\Gamma^t_{\ \theta\theta}(r),\\
\mathcal{B}(r)&=\frac{(2 c-k)\Gamma^t_{\ \theta\theta}(r)[1+c\Gamma^t_{\ \theta\theta}(r)]}{\Gamma^r_{\ \theta\theta}(r)},\\
\mathcal{C}(r)&= -c(2c-k)\Gamma^r_{\ \theta\theta}(r),\\
\mathcal{D}(r)&=c+c(2c-k)\Gamma^t_{\ \theta\theta}(r),\\
\mathcal{E}(r)&=-\frac{1+c\Gamma^t_{\ \theta\theta}(r)}{\Gamma^r_{\ \theta\theta}(r)},
\end{align}
\end{subequations}
 $c,k$ are real constant, and everything depends on the functions $\Gamma^t_{\ rr}(r),\Gamma^t_{\ \theta\theta}(r),\Gamma^r_{\ rr}(r),\Gamma^r_{\ \theta\theta}(r)$ with $\Gamma^r_{\ \theta\theta}(r)\neq0$. In addition, we have further constraints \cite{DAmbrosio2022}
\begin{subequations}
\begin{align}
\Gamma^t_{\ \theta\theta}{}'(r)&=-\frac{\Gamma^t_{\ \theta\theta}(r)}{\Gamma^r_{\ \theta\theta}(r)}\biggr{[}1+\Gamma^t_{\ \theta\theta}(r)(3c-k\notag\\
&+(2c-k)\Gamma^t_{\ \theta\theta}(r))\biggr{]}-\Gamma^t_{\ rr}(r)\Gamma^r_{\ \theta\theta}(r),\\
\Gamma^r_{\ \theta\theta}{}'(r)&=-1-c\Gamma^t_{\ \theta\theta}(r)\biggr{[}2+(2c-k)\Gamma^t_{\ \theta\theta}(r)\biggr{]}\notag\\
&-\Gamma^t_{\ rr}(r)\Gamma^r_{\ \theta\theta}(r),
\end{align}
\end{subequations}
where $\Gamma^t_{\ rr}{}'(r)$ and $\Gamma^r_{\ rr}{}'(r)$ cannot be expressed in terms of the other components. There are additional restrictions depending on the fact that the field equations must be diagonal and its $tr$ component must vanish. This leads to distinguish two cases (see Table 3 in Ref. \cite{DAmbrosio2022})
\begin{subequations} \label{eq:CASES}
\begin{align}
&\mbox{if}\ c\neq0\ \mbox{and}\ k\neq2c \notag\\ 
&\begin{cases}
\Gamma^t_{\ \theta\theta}(r)&=\frac{k}{2c(2c-k)},\\
\Gamma^t_{\ rr}(r)&=-\frac{k(8c^2+2ck-k^2)}{8c^2(2c-k)^2(\Gamma^r_{\ \theta\theta}(r))^2};
\end{cases}\label{eq:case1}\\
&\mbox{if}\ c=0\ \mbox{and}\ k=0\notag\\ 
&\begin{cases}
\Gamma^t_{\ rr}(r)&=-\frac{\Gamma^t_{\ \theta\theta}(r)}{(\Gamma^r_{\ \theta\theta}(r))^2}.
\end{cases}\label{eq:case2}
\end{align}
\end{subequations}

Depending on the chosen expression of the affine connection, the ensuing calculations can be carried out and entail different outcomes. Therefore, we can distinguish two independent scenarios.
\begin{itemize}
% 1 CASE
\item \emph{Case \eqref{eq:case1}.} The terms enclosed in the square brackets of Eq. \eqref{eq:inequality_1} reads as
\begin{align} 
&P^r_{\ rr}\left(1-\frac{b(r)}{r}\right)+\frac{P^r_{\ tt}}{e^{2\Phi(r)}}\notag\\
&=\frac{b(r)-r}{8 r^2} \left[\frac{r (4 c-k) \left(8 c^2-6 c k+k\right)}{c (k-2 c)^2 \Gamma^r_{\ \theta \theta}(r)}+8\right]\notag\\
&\quad+(c-1) c \Gamma^r_{\ \theta \theta}(r) e^{-2 \Phi (r)}.
\end{align} 
Therefore, Eq. \eqref{eq:inequality_1} becomes
\begin{align} \label{eq:inequality_21}
&-2\ddot{f}(Q)Q'(r)\Biggr{[}\frac{b(r)-r}{8 r^2} \left(\frac{r (4 c-k) \left(8 c^2-6 c k+k\right)}{c (k-2 c)^2 \Gamma^r_{\ \theta \theta}(r)}+8\right)\notag\\
&+(c-1) c \Gamma^r_{\ \theta \theta}(r) e^{-2 \Phi (r)}\Biggr{]}\ge0.
\end{align}
Depending on the sign of the terms in the square brackets of Eq. \eqref{eq:inequality_21}, we can also infer the sign of $\ddot{f}(Q)Q'(r)$. Nevertheless, even if we evaluate Eq. \eqref{eq:inequality_21} at the WH throat, we still do not gather helpful information. Analysing the terms in the square brackets with respect to $\Gamma^r_{\ \theta \theta}(r)$, we have a second order algebraic equation admitting two real roots
\begin{subequations}
\begin{align}
\mathcal{S}_+&=\frac{e^{\Phi (r)} \left[2 (r-b(r)) (2 c-k) e^{\Phi (r)}+\sqrt{2} \sqrt{\Delta}\right]}{4 (c-1) c r^2 (2 c-k)}, \label{eq:Splus}\\
\mathcal{S}_-&=\frac{e^{\Phi (r)} \left[2 (r-b(r)) (2 c-k) e^{\Phi (r)}-\sqrt{2} \sqrt{\Delta}\right]}{4 (c-1) c r^2 (2 c-k)}
,\label{eq:Sminus}\\
\Delta&=(r-b(r)) \left[2 (r-b(r)) (k-2 c)^2 e^{2 \Phi (r)}\right.\notag\\
&\left.+(c-1) r^3 (4 c-k) \left(8 c^2-6 c k+k\right)\right]. \label{eq:Delta}
\end{align}
\end{subequations}
Emulating the previous gravity frameworks, we may assume $\ddot{f}(Q)Q'(r)\ge0$, from which we obtain
\begin{equation} \label{eq:inequality_21bis}
\mathcal{S}_-\le \Gamma^r_{\ \theta \theta}(r)\le\mathcal{S}_+.
\end{equation}
Now, imposing that $\Delta\ge0$, we get
\begin{align} \label{eq:inequality_31}
b(r)=r,\ \mbox{or}\ b(r)\le\bar{b}(r)\le r,
\end{align}
where 
\begin{align}
 \bar{b}(r)&=\frac{r^3(c-1) (4 c-k) \left(8 c^2-6 c k+k\right)}{2e^{2 \Phi (r)} (k-2 c)^2}+r.  \label{eq:barb}
\end{align}
The condition $\bar{b}(r)\le r $ implies
\begin{equation} \label{eq:inequality_41}
(c-1) (4 c-k) \left(8 c^2-6 c k+k\right)\leq 0.
\end{equation}
Analysing also the NEC \eqref{eq:NEC_phi}, where we use
\begin{align}
\frac{P^r_{\ tt}}{g_{tt}}+\frac{P^r_{\ \varphi\varphi}}{g_{\varphi\varphi}}&=\frac{r [r-b(r)] \Phi '(r)+b(r)-\Gamma^r_{\ \theta \theta}(r)-r}{2 r^2}\notag\\
&+(c-1) c \Gamma^r_{\ \theta \theta}(r) e^{-2 \Phi (r)},
\end{align}
we finally obtain
\begin{align}
-2\ddot{f}(Q)Q'(r)&\biggr{[}\frac{r (r-b(r)) \Phi '(r)+b(r)-\Gamma^r_{\ \theta \theta}(r)-r}{2 r^2}\notag\\
&+(c-1) c \Gamma^r_{\ \theta \theta}(r) e^{-2 \Phi (r)}\biggr{]}\ge0.
\end{align}
This allows to get 
\begin{align} \label{eq:inequality_51}
\Phi'(r)\le\frac{r-b(r)+\Gamma^r_{\ \theta \theta}(r)\left[1-\frac{2 (c-1) c r^2}{ e^{2 \Phi (r)}}\right]}{r (r-b(r))}.
\end{align}
% 2 CASE
\item \emph{Case \eqref{eq:case2}.} The terms enclosed in the square brackets of Eq. \eqref{eq:inequality_1} now reads as
\begin{align} 
&\frac{P^r_{\ tt}}{e^{2\Phi(r)}}+P^r_{\ rr}\left(1-\frac{b(r)}{r}\right)\notag\\
=&-(r-b(r)) \left[\frac{2 \Gamma^r_{\ \theta \theta}(r)+r \biggr{(}\Gamma^t_{\ \theta \theta}(r)+2\biggr{)}}{2 r^2 \Gamma^r_{\ \theta \theta}(r)}\right].
\end{align} 
Therefore, Eq. \eqref{eq:inequality_1} becomes
\begin{align} \label{eq:inequality_22}
&\ddot{f}(Q)Q'(r)\biggr{(}r-b(r)\biggr{)} \notag\\
&\times\left[\frac{2 \Gamma^r_{\ \theta \theta}(r)+r \biggr{(}\Gamma^t_{\ \theta \theta}(r)+2\biggr{)}}{2 r^2 \Gamma^r_{\ \theta \theta}(r)}\right]\ge0.
\end{align}
Also in this case, the inequality strongly depends on the sign of the terms inside the square brackets, and even if we evaluate this expression at the WH throat, we do not have more information. Therefore, as done in the previous case, we assume that $\ddot{f}(Q)Q'(r)\ge0$, which implies
\begin{align} \label{eq:inequality_12}
\frac{2 \Gamma^r_{\theta \theta}(r)+r (\Gamma^t_{\theta \theta}(r)+2)}{\Gamma^r_{\theta \theta}(r)}\ge0.
\end{align}

Analysing also the NEC \eqref{eq:NEC_phi}, where we use
\begin{align}
&\frac{P^r_{\ tt}}{g_{tt}}+\frac{P^r_{\ \varphi\varphi}}{g_{\varphi\varphi}}\notag\\
=&\frac{(r-b(r)) \left(r \Phi '(r)-1\right)-\Gamma^r_{\ \theta \theta}(r)}{2 r^2},
\end{align}
we finally obtain
\begin{align}
-2Q'(r)\ddot{f}(Q)&\biggr{[}\frac{(r-b(r)) \left(r \Phi '(r)-1\right)-\Gamma^r_{\ \theta \theta}(r)}{2 r^2}\biggr{]}\ge0.
\end{align}
This entails 
\begin{align} \label{eq:inequality_22}
\Phi'(r)\le\frac{r-b(r)+\Gamma^r_{ \ \theta \theta}(r)}{r(r-b(r))}.
\end{align}
\end{itemize}
Since in this theory metric and affine structures are separated, the inequalities become in general more tangled with respect to the $f(R)$ and $f(T)$ theories. The assumption $Q(r)>0$ is a complicate function of the affine connections and metric components in both cases \eqref{eq:case1} and \eqref{eq:case2}. Thus, from this last  condition it is not possible to extract a simple inequality. Therefore, the general constraints we have derived are: (1) for the case \eqref{eq:case1}, we have Eqs. \eqref{eq:inequality_21bis}, \eqref{eq:inequality_31}, \eqref{eq:inequality_41}, \eqref{eq:inequality_51}, and $Q(r)>0$; (2) for the case \eqref{eq:case2}, we have Eqs. \eqref{eq:inequality_12}, \eqref{eq:inequality_22}, and $Q(r)>0$ 

\section{Discussion and Conclusions}
\label{sec:end}
In this paper, we have analysed static and spherically symmetric WH solutions, parametrized by the redshift $\Phi(r)$ and shape $b(r)$ functions in $f(R)$, $f(T)$, and $f(Q)$ theories in the vacuum. In other words, we exploited only gravity without resorting to any kind of exotic matter stress-energy tensor. As preliminary requirements we impose $b(r)\le r$ and $b'(r)<b(r)/r$ (i.e., flaring out condition), as well as other regularity conditions for avoiding the appearance of singularities\footnote{The Schwarzschild solution satisfies both NEC and flaring out condition, but it is not a WH, because it posses the physical singularity in $r=0$. This example shows that to have a WH solution, we must be sure to satisfy all the requirements.}, see Sec. \ref{sec:ss_wormholes}. After, we use the (gravitational) curvature fluid energy-momentum tensor $T^{\rm (curv)}_{\alpha\beta}$ as the source to make the WH traversable, and then we disclose the calculations for the NEC to provide further constraints on the gravity theory itself and on the metric components (see Table \ref{tab:Table1}).
\renewcommand{\arraystretch}{1.6}
\begin{table}[th!]
\begin{center}
\caption{\label{tab:Table1} Summary of the constraints on the WH solutions determined in different gravity frameworks, where the further degrees of freedom of the extended gravitational theories can be the source to build up WH topological structures.}	
%\normalsize
\footnotesize
\vspace{0.3cm}
\begin{tabular}{|@{} c @{}|@{} c @{}|} 
\ChangeRT{1pt}
\quad {\bf THEORY} \quad  &\quad {\bf CONSTRAINTS} \quad \\
\hline
& $R'(r)\ddot{f}(R)\ge0$ \\
& $\dddot{f}(R)\le -\frac{\ddot{f}(R)}{R'(r)}\left[\frac{b(r)-b'(r)r}{2r(r-b(r))}-\Phi'(r)+\frac{R''(r)}{R'(r)}\right]$\\
$f(R)$& $\Phi_1(r)<\Phi'(r)<\Phi_2(r)\le \frac{1}{r}$\footnote{The functions $\Phi_1(r)$ and $\Phi_2(r)$ are given in Eq. \eqref{eq:SYMBOLS}.}\\
& $\Phi ''(r)<\frac{1}{r (r-b(r))}$\\
& $b(r)\le r\left[\frac{r^2 \Phi ''(r)+3}{r^2 \Phi ''(r)+4}\right]$\\
\hline
$f(T)$ & $\ddot{f}(T) T'(r)\ge0$ \\
 & $\Phi '(r)< -\frac{1}{2 r}$ \\
 \hline
& $\ddot{f}(Q)Q'(r)\ge0\ \mbox{(assumption)}$\\
& $\mathcal{S}_- \le \Gamma^r_{\ \theta \theta}(r) \le \mathcal{S}_+$ \footnote{The expressions of $\mathcal{S}_-,\mathcal{S}_+$ are reported in Eqs. \eqref{eq:Splus} and \eqref{eq:Sminus}.}\\
$f(Q)$ & $b(r)\le \bar{b}(r)\le r$ \footnote{The expression of $\bar{b}(r)$ can be found in Eq. \eqref{eq:barb}.}\\
CASE \eqref{eq:case1} & $(c-1) (4 c-k) \left(8 c^2-6 c k+k\right)\leq 0$\\
& $\Phi'(r)\le\frac{r-b(r)+\Gamma^r_{\ \theta \theta}(r)\left[1-\frac{2 (c-1) c r^2}{ e^{2 \Phi (r)}}\right]}{r (r-b(r))}$\\
& $Q(r)>0$\\
 \hline
  & $\ddot{f}(Q)Q'(r)\ge0\ \mbox{(assumption)}$\\
$f(Q)$ & $\frac{2 \Gamma^r_{\theta \theta}(r)+r (\Gamma^t_{\theta \theta}(r)+2)}{\Gamma^r_{\theta \theta}(r)}\ge0$ \\
CASE \eqref{eq:case2}& $\Phi'(r)\le\frac{r-b(r)+\Gamma^r_{ \ \theta \theta}(r)}{r(r-b(r))}$\\
& $Q(r)>0$\\
\ChangeRT{1pt}
\end{tabular}
\end{center}
\end{table}

Therefore, applying the NEC conditions, we have found useful inequalities in $f(R)$ and $f(T)$ theories, which permitted to fix the traversability issue. Nevertheless, in $f(Q)$ theory the situation is more tangled due to the independence of the metric tensor and the affine connections. This is the main reason, which leads to distinguish two cases depending on the choice of the affine connections. Having further information on the functional form of the affine connections, it is possible to infer tighter constraints on the WH solutions. 

As an application, let us consider the following general models $f_1(X)=X^{1+\varepsilon}$ and $f_2(X)=X+\varepsilon X \log X$, where $|\varepsilon|\ll1$. The task is to fix the sign of $\varepsilon$ in order to obtain a WH solution sourced only by gravity. As shown in Table \ref{tab:Table1}, we have $\ddot{f}(X)x'\ge0$ in all the considered gravity frameworks. Assuming that $X\sim r^{-\alpha}>0$ with $\alpha >0$, this leads to $X\sim r^{-\alpha-1}<0$. Therefore, we obtain
\begin{subequations}
\begin{align} 
\ddot{f}_1(X)X'&=\varepsilon(1+\varepsilon)X^{\varepsilon-1}X'\ge0\ \ \Rightarrow\ \ \varepsilon\le0,\\
\ddot{f}_2(X)X'&=\varepsilon\frac{X'}{X}\ge0\hspace{2.03cm} \Rightarrow\ \ \varepsilon\le0.
\end{align} 
\end{subequations}
This example is very significant, because it demonstrates that although we have not yet explicitly determined the WH solution, thanks to the derived inequalities it is possible to select at least the sign of $\epsilon$. This result is thus important for determining the underlying gravity theory.

Another fundamental issue relies upon the stability of the WH solution. This condition does not provide inequalities as done for the traversability, because the involved calculations are more cumbersome. A WH solution sourced only by gravity can be conceived in a different manner with respect to that supported by an exotic matter fluid. Indeed, $T_{\mu\nu}^{\rm (curv)}$ can be seen as a topological and geometric fluid endowed with \qm{a rigid structure}. For these reasons, we can consider a kind of \emph{hydrodynamic stability}, which implies to impose the following conditions along the radial and tangential directions, respectively. They are given by \cite{Morris1988}
\begin{subequations}
\begin{align}
\left(\frac{\partial p_r(r)}{\partial \rho(r)}\right)_{r=b_0}=\left(\frac{p'_r(r)}{\rho'(r)}\right)_{r=b_0}=0,\\
\left(\frac{\partial p_t(r)}{\partial \rho(r)}\right)_{r=b_0}=\left(\frac{p'_t(r)}{\rho'(r)}\right)_{r=b_0}=0,
\end{align}
\end{subequations}
and employing Eqs. \eqref{eq:components_curvature_tensor},  we have
\begin{subequations}
\begin{align} 
&\left(\frac{p'_r(r)}{\rho'(r)}\right)_{r=b_0}\notag\\
&=\frac{T_{rr}^{\rm (curv)}(b_0) e^{2 \Phi (b_0)} \left(b'(b_0)-1\right)}{b_0\left(T_{tt}^{\rm (curv)}{}'(b_0)-2 T_{tt}^{\rm (curv)}(b_0) \Phi '(b_0)\right)}=0,\label{eq:stability_rad}\\
&\left(\frac{p'_t(r)}{\rho'(r)}\right)_{r=b_0}\notag\\
&=\frac{e^{2 \Phi (b_0)} \left(2 T^{\rm (curv)}_{\varphi \varphi}(b_0)-r T^{\rm (curv)}_{\varphi \varphi}{}'(b_0)\right)}{b_0^3 \left(T^{\rm (curv)}_{tt}{}'(b_0)-2 T_{tt}^{\rm (curv)}(b_0) \Phi '(b_0)\right)}=0.\label{eq:stability_tan}
\end{align} 
\end{subequations}
It could be also possible to calculate an {\it average hydrodynamic stability} by defining the following average pressure $p(r)=\frac{1}{3}[p_r(r)+2p_t(r)]$. In this way, we obtain
\begin{align} \label{eq:stability_av}
&\left(\frac{\partial p(r)}{\partial \rho(r)}\right)_{r=b_0}=\left(\frac{p'(r)}{\rho'(r)}\right)_{r=b_0}\notag\\
&=\frac{e^{2 \Phi (b_0)}}{3 b_0^3}\left[ \frac{b_0^2 T_{rr}^{\rm (curv)}(b_0) \left(b'(b_0)-1\right)-2 b_0 T^{\rm (curv)}_{\varphi \varphi}{}'(b_0)}{T^{\rm (curv)}_{tt}{}'(b_0)-2 T_{tt}(b_0) \Phi '(b_0)}\right.\notag\\
&\left.+\frac{4 T^{\rm (curv)}_{\varphi \varphi}(b_0)}{T^{\rm (curv)}_{tt}{}'(b_0)-2 T_{tt}(b_0) \Phi '(b_0)}\right]=0.
\end{align}
The computations of such terms gives not simple equations, but must be disclosed once one posses an explicit WH solution framed in a precise gravity framework.  However, to be more precise, the stability requirement demands to check the behavior of the solution under time-dependent perturbations. This is of course a more complex problem, which must be verified once we determine a solution. Nevertheless, the strategy based on first checking the hydrodynamic stability could give already a hint for fulfilling the more general claim.

As final remark, we can say that our approach constitutes general indications, represented by Table \ref{tab:Table1}, for constructing WH solutions supported only by gravity in $f(R)$, $f(T)$, and $f(Q)$ theories. This strategy has never been investigated in the literature, to the best of our knowledge. The above calculations and results allow not only to inquire the existence of viable WH solutions, but also to figure out possible links among different gravity theories. In a forthcoming paper, possible astrophysical applications will be discussed.

\section*{Acknowledgements}
This paper is based upon work from COST Action CA21136: {\it Addressing observational tensions in cosmology with systematics and fundamental physics} (CosmoVerse) supported by COST (European Cooperation in Science and Technology). Authors acknowledge the Istituto Nazionale di Fisica Nucleare (iniziative specifiche QGSKY, TEONGRAV, and MOONLIGHT2) and the Gruppo Nazionale di Fisica Matematica (Istituto Nazionale di Alta Matematica) for the support.  

\bibliography{references}

\end{document}